\documentclass[preprint,5p,sort&compress]{elsarticle}

\usepackage{amsmath,amssymb}
\usepackage{graphics,graphicx}
\usepackage{dcolumn}
\usepackage{bm}
\usepackage{epsfig}
\usepackage[usenames]{color}
\usepackage{hyperref} 
\usepackage{mathbbol}
\usepackage{epstopdf}
\usepackage{simplewick}
\usepackage[utf8]{inputenc} 
\usepackage{slashed}
\usepackage{stackrel}

\allowdisplaybreaks

\def\eq#1{{Eq.~(\ref{#1})}}

\newcommand{\ben}{\begin{eqnarray*}}
\newcommand{\een}{\end{eqnarray*}}

\newcommand{\as}{\alpha_s}

\newcommand{\dhd}{{\textstyle d}
\lower.03ex\hbox{\kern-0.38em$^{\scriptstyle-}$}\kern-0.05em{}}
\newcommand{\dbar}{{\textstyle \delta}
\lower.03ex\hbox{\kern-0.38em$^{\scriptstyle-}$}\kern-0.05em{}}

\begin{document}

\title{Small-$x$ Asymptotics of the Quark Helicity Distribution: Analytic Results }

\author[YK]{Yuri V. Kovchegov} 
         \ead{kovchegov.1@osu.edu}
         \address[YK]{Department of Physics, The Ohio State
           University, Columbus, OH 43210, USA}
\author[DP]{Daniel Pitonyak}
	\ead{dap67@psu.edu}
	\address[DP]{Division of Science, Penn State University-Berks, Reading, PA 19610, USA}

\author[MS]{Matthew D. Sievert}
        \ead{sievertmd@lanl.gov}
	\address[MS]{Theoretical Division, Los Alamos National Laboratory, Los Alamos, NM 87545, USA}

\begin{abstract}
  In this Letter, we analytically solve the evolution equations for
  the small-$x$ asymptotic behavior of the (flavor singlet) quark
  helicity distribution in the large-$N_c$ limit.  These evolution
  equations form a set of coupled integro-differential equations,
  which previously could only be solved numerically.  This approximate
  numerical solution, however, revealed simplifying properties of the
  small-$x$ asymptotics, which we exploit here to obtain an analytic
  solution.  We find that the small-$x$ power-law tail of the quark
  helicity distribution scales as $\Delta q^S (x, Q^2) \sim
  \left(\tfrac{1}{x} \right)^{\alpha_h}$ with $\alpha_h =
  \tfrac{4}{\sqrt{3}} \sqrt{\tfrac{\alpha_s N_c}{2\pi}}$, in excellent
  agreement with the numerical estimate $\alpha_h \approx
  2.31\sqrt{\tfrac{\alpha_s N_c}{2\pi}}$ obtained previously.  We then
  verify this solution by cross-checking the predicted scaling
  behavior of the auxiliary ``neighbor dipole amplitude" against the
  numerics, again finding excellent agreement.
\end{abstract}

\begin{keyword}
\PACS 12.38.-t \sep 12.38.Bx \sep 12.38.Cy
\end{keyword}

\maketitle


\section{Introduction}

The small-$x$ power-law behavior of parton distribution functions
(PDFs) and hadronic structure functions at small Bjorken $x$ is
governed by quantum evolution equations which resum large logarithms
of $\tfrac{1}{x}$.  The most familiar of these is the linear
Balitsky-Fadin-Kuraev-Lipatov (BFKL) equation \cite{Kuraev:1977fs,
  Balitsky:1978ic} for the unpolarized structure functions $F_1$ and
$F_2$ along with the quark and gluon PDFs at small $x$, which resums
the single-logarithmic parameter $\alpha_s \ln\tfrac{1}{x} \sim 1$
(with $\as$ the strong coupling constant).  The result of this
resummation is a power-law growth at small $x$ given by $F_1 (x,Q^2)
\sim q(x,Q^2) \sim \left(\tfrac{1}{x} \right)^{\alpha_P}$, with the
leading-order (LO) exponent $\alpha_p = 1 + \tfrac{4 \alpha_s
  N_c}{\pi} \ln2$ known as the perturbative ``Pomeron intercept'' in the
terminology of Regge theory.  Here $N_c$ is the number of colors.

The analogous small-$x$ asymptotic behavior of the helicity PDFs
$\Delta f (x, Q^2)$ and the polarized structure function $g_1 (x,
Q^2)$ has received much less attention than the unpolarized
case. Early studies emphasized the role of exchanging polarized quarks
\cite{Kirschner:1983di, Kirschner:1985cb, Kirschner:1994vc,
  Kirschner:1994rq, Griffiths:1999dj, Itakura:2003jp} (the ``Reggeon''
in Regge theory), with important progress on the full polarized
evolution made by Bartels, Ermolaev, and Ryskin \cite{Bartels:1995iu,
  Bartels:1996wc}.  Recently, we have derived the small-$x$ evolution
equations for the quark helicity PDFs $\Delta q^S (x, Q^2)$ and the
polarized structure function $g_1 (x, Q^2)$ \cite{Kovchegov:2015pbl,
  Kovchegov:2016zex} in the modern language of the dipole model.  (In
this Letter, we restrict our discussion to the flavor-singlet quark
helicity distribution; for the non-singlet quark helicity
distribution, see \cite{Kovchegov:2016zex}.)  These helicity evolution
equations, like the perturbative Reggeon evolution equations, resum
the double-logarithmic parameter $\alpha_s \ln^2 \tfrac{1}{x} \sim 1$;
they couple to both polarized quark and gluon exchange, and, in this
respect, differ from the gluon-only unpolarized LO BFKL equation.

In general, the helicity evolution equations derived in
\cite{Kovchegov:2015pbl, Kovchegov:2016zex} form an infinite tower of
operator equations analogous to the Balitsky hierarchy
\cite{Balitsky:1995ub, Balitsky:1998ya} for the unpolarized small-$x$
evolution
\cite{Balitsky:1996ub,Balitsky:1998ya,Kovchegov:1999yj,Kovchegov:1999ua,Jalilian-Marian:1997dw,Jalilian-Marian:1997gr,Iancu:2001ad,Iancu:2000hn}.
In both cases, the operator hierarchy closes in the large-$N_c$ limit
\cite{Balitsky:1996ub,Balitsky:1998ya,Kovchegov:1999yj,Kovchegov:1999ua}.
For helicity evolution, this still yields a pair of coupled
integro-differential equations for the ``polarized dipole amplitude''
$G(x_{10}^2 , z )$ and the auxiliary ``neighbor dipole amplitude''
$\Gamma (x_{10}^2 , x_{21}^2 , z)$ that must be solved to determine
the power-law behavior at small $x$. (Here $x_{ij}$'s denote
transverse sizes of dipoles and $z$ is the softest longitudinal
momentum fraction between the quark and antiquark in the dipole.) This
asymptotic behavior of the polarized dipole $G(x_{10}^2 , z) \sim (z
s)^{\alpha_h}$ determines the corresponding small-$x$ asymptotics of
the helicity PDFs and the polarized structure function: $\Delta q^S
(x, Q^2) \sim g_1 (x, Q^2) \sim \left(\tfrac{1}{x}
\right)^{\alpha_h}$, where we refer to the exponent $\alpha_h$ as the
``helicity intercept'' in analogy to the Pomeron intercept.

In \cite{Kovchegov:2016weo}, we solved the large-$N_c$ helicity
evolution equations for $\alpha_h$ numerically, obtaining $\alpha_h
\approx 2.31 \sqrt{\tfrac{\alpha_s N_c}{2\pi}}$. We also found that
such an intercept could lead to a significant enhancement of the
contribution from the quark spin $\Delta\Sigma$ to the proton
spin~\cite{Kovchegov:2016weo}, which is not ruled out by current
experimental data~\cite{Nocera:2016zyg}.  In the following Sections,
we use an emergent scaling feature of this numerical solution, namely,
that $G$ depends only on a single combination of its arguments and not
on each independently, to derive an analytic expression for
$\alpha_h$.

\section{Solution of the large-$N_c$ equations}

In standard coordinates, the large-$N_c$ helicity evolution equations
read \cite{Kovchegov:2015pbl, Kovchegov:2016zex}
\begin{subequations} \label{e:helevol}
\begin{align}
  G(x_{10}^2 , z) &= G^{(0)} (x_{10}^2 , z) + \frac{\alpha_s
    N_c}{2\pi} \int\limits_{\tfrac{1}{x_{10}^2 s}}^{z} \frac{d z'}{z'}
  \int\limits_{\tfrac{1}{z' s}}^{x_{10}^2} \frac{d x_{21}^2}{x_{21}^2}
\notag \\ &\times
\Big[ \Gamma(x_{10}^2 , x_{21}^2 , z') + 3 G(x_{21}^2 , z') \Big] ,
\\ \notag \\
\Gamma(x_{10}^2 , x_{21}^2 , z') &= G^{(0)}(x_{10}^2 , z') +
\frac{\alpha_s N_c}{2\pi} \int\limits_{\tfrac{1}{x_{10}^2 s}}^{z'}
\frac{dz''}{z''}
\notag \\ & \hspace{-1.25cm} \times
\hspace{-0.75cm} \int\limits_{\tfrac{1}{z'' s}}^{\mbox{min}\left\{
    x_{10}^2 \, , \, x_{21}^2 \tfrac{z'}{z''} \right\}}
\hspace{-0.75cm} \frac{dx_{32}^2}{x_{32}^2} \Big[ \Gamma(x_{10}^2 ,
x_{32}^2 , z'') + 3 G(x_{32}^2 , z'') \Big] ,
\end{align}
\end{subequations}
where $x_{10} , x_{21} , x_{32}$ are the transverse sizes of various
dipoles and $z , z' , z''$ are longitudinal momentum fractions of the
softest (anti-)quarks in the dipoles.  Following
\cite{Kovchegov:2016weo}, it is convenient to introduce the scaled
logarithmic variables
\begin{subequations} \label{e:logunits}
\begin{align}
  \eta & \equiv \sqrt{\frac{\alpha_s N_c}{2\pi}} \ln\frac{z
    s}{\Lambda^2} , & \hspace{0.5cm} s_{10} & \equiv
  \sqrt{\frac{\alpha_s N_c}{2\pi}} \ln\frac{1}{x_{10}^2 \Lambda^2} ,
\\
\eta' & \equiv \sqrt{\frac{\alpha_s N_c}{2\pi}} \ln\frac{z' s}{\Lambda^2} ,
& \hspace{0.5cm}
s_{21} & \equiv \sqrt{\frac{\alpha_s N_c}{2\pi}} \ln\frac{1}{x_{21}^2 \Lambda^2} ,
\\
\eta'' & \equiv \sqrt{\frac{\alpha_s N_c}{2\pi}} \ln\frac{z'' s}{\Lambda^2} ,
& \hspace{0.5cm}
s_{32} & \equiv \sqrt{\frac{\alpha_s N_c}{2\pi}} \ln\frac{1}{x_{32}^2 \Lambda^2} ,
\end{align}
\end{subequations}
where $\Lambda$ is an IR momentum cutoff and $s$ is the
center-of-mass-energy squared at which the helicity PDF is measured.
In terms of these rescaled variables, the large-$N_c$ helicity
evolution equations are
\begin{subequations}\label{GGeqns}
\begin{align} \label{e:G_redef} 
  & G (s_{10}, \eta) = G^{(0)} (s_{10}, \eta) +
  \int\limits_{s_{10}}^\eta d \eta' \int\limits_{s_{10}}^{\eta'} d
  s_{21} \,\Big[ \Gamma (s_{10}, s_{21}, \eta')
\notag \\ & \hspace{1.4cm}
+ 3 \, G (s_{21}, \eta') \Big],   
\\ \notag \\ \label{e:Gam_redef}
& \Gamma (s_{10}, s_{21}, \eta') = G^{(0)} (s_{10}, \eta') +
\int\limits_{s_{10}}^{\eta'} d \eta'' \hspace{-0.4cm}
\int\limits_{\mbox{max} \left\{ s_{10}, s_{21} + \eta'' - \eta'
  \right\}}^{\eta''} \hspace*{-1cm} d s_{32}
\notag \\ & \hspace{2.1cm} \times
\left[ \Gamma (s_{10}, s_{32}, \eta'') + 3 \, G (s_{32}, \eta'') \right]. 
\end{align}
\end{subequations}

\begin{figure}[ht]
\includegraphics[width=0.48\textwidth]{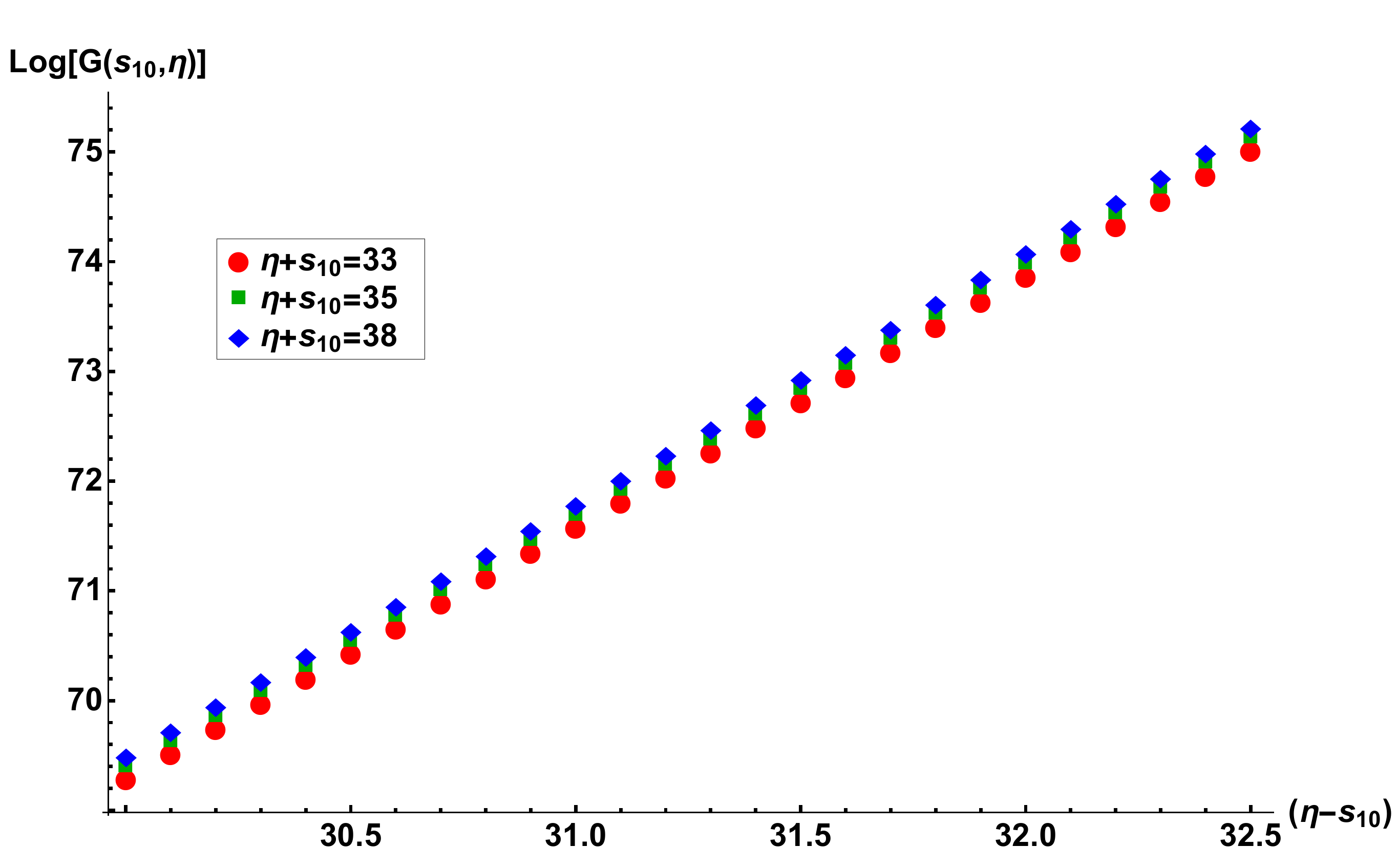}\\
\includegraphics[width=0.48\textwidth]{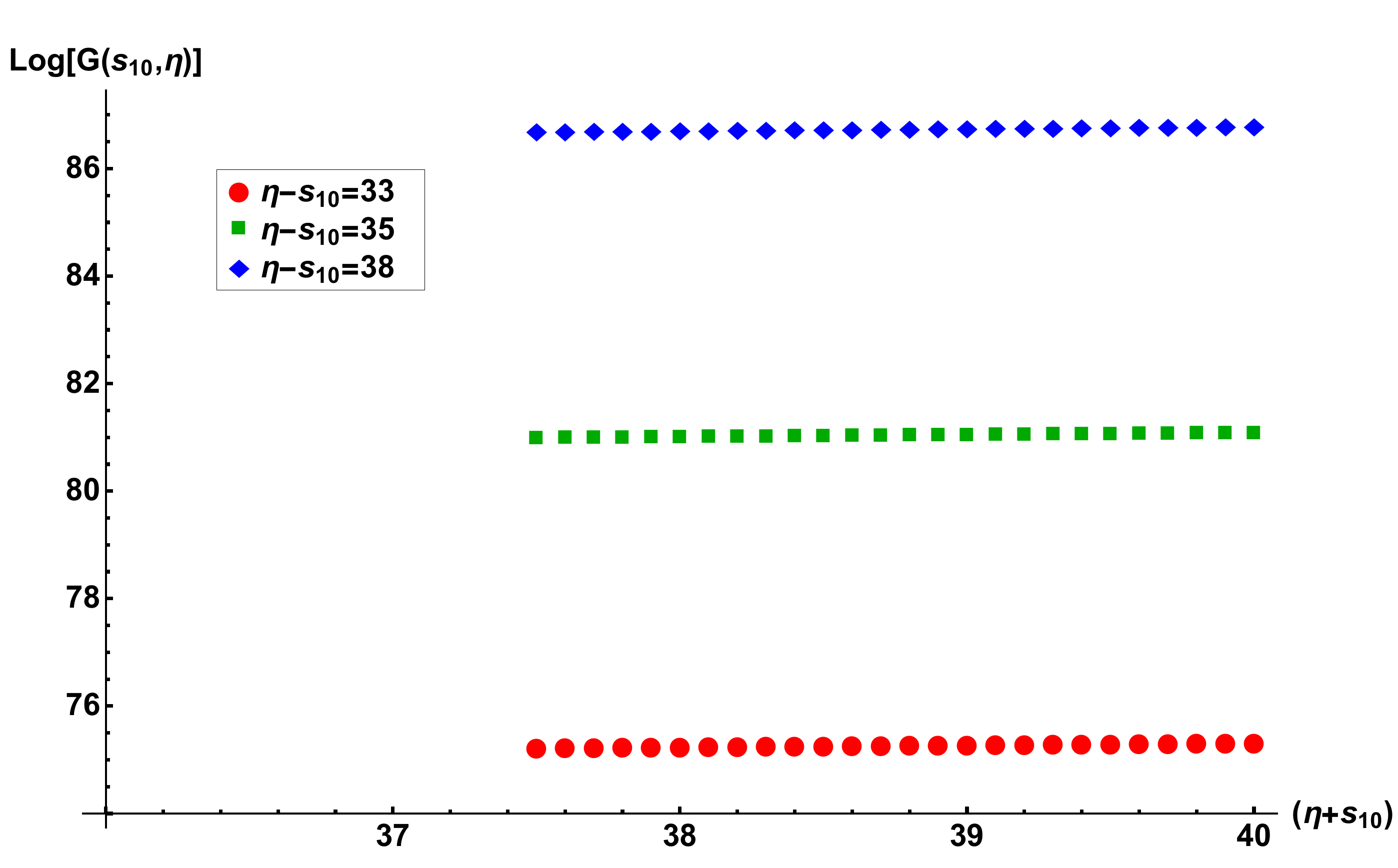}
\caption{ \label{f:num1} Numerical solution of the scaled equations
  \eqref{GGeqns} as a function of $\eta-s_{10}$ for fixed
  $\eta+s_{10}$ (top panel) and as a function of $\eta+s_{10}$ for
  fixed $\eta-s_{10}$ (bottom panel).  The grid parameters are
  $\eta_{max} = 40 , \Delta\eta = 0.05$.  One clearly sees that $\ln
  G$ has a linear dependence on $\eta-s_{10}$ and is independent of
  $\eta+s_{10}$.  }
\end{figure}

In the numerical solution of \cite{Kovchegov:2016weo}, two important
features were observed in the asymptotic limit: (a) a negligible
dependence on the choice of initial conditions $G^{(0)}$ and (b) the
dependence of $G$ only on the combination $\zeta \equiv \eta -
s_{10}$, rather than on $\eta$ and $s_{10}$ separately (see
Fig.~\ref{f:num1}).  Let us therefore (a) trivially fix the initial
conditions to $G^{(0)} = 1$ and (b) assume the $\eta - s_{10}$ scaling
from the outset:
\begin{subequations}\label{scaling}
\begin{align}
G (s_{10}, \eta) &= G (\eta - s_{10}), 
\\
\Gamma (s_{10}, s_{21}, \eta') &= \Gamma (\eta' - s_{10}, \eta' - s_{21}).
\end{align}
\end{subequations}
Then Eqs.~\eqref{GGeqns}  become
\begin{subequations}\label{GGeqns2}
\begin{align} \label{e:G_redef2}
  G (\zeta) & = 1 + \int\limits_{0}^\zeta d \xi \int\limits_{0}^{\xi}
  d \xi' \,\left[ \Gamma (\xi, \xi') + 3 \, G (\xi') \right] ,
\\ 
\Gamma (\zeta, \zeta') & = 1 + \int\limits_{0}^{\zeta'} d \xi
\int\limits_{0}^{\xi} d \xi' \,\left[ \Gamma (\xi, \xi') + 3 \, G
  (\xi') \right]
\notag \\ &+
\int\limits_{\zeta'}^{\zeta} d \xi  \, \int\limits_{0}^{\zeta'} d \xi' 
\left[ \Gamma (\xi, \xi') + 3 \, G (\xi')  \right] 
\notag \\ \label{e:Gam_redef2} & =
G (\zeta') + \int\limits_{\zeta'}^{\zeta} d \xi \,
\int\limits_{0}^{\zeta'} d \xi' \left[ \Gamma (\xi, \xi') + 3 \, G
  (\xi') \right] ,
\end{align}
\end{subequations}
with the boundary conditions
\begin{align}\label{init}
G (0) =1, \ \ \ \Gamma (\zeta', \zeta') = G (\zeta').
\end{align}
Since $s_{10} \le s_{21}$, we consider $\Gamma(\zeta, \zeta')$ only in
the range $\zeta > \zeta'$.

To solve \eqref{GGeqns2}, we first differentiate, obtaining
\begin{subequations}\label{GGeqns3}
\begin{align} 
  \partial_\zeta G (\zeta) & = \int\limits_{0}^{\zeta} d \xi'  \,\left[ \Gamma (\zeta, \xi') + 3 \, G (\xi') \right],  \label{e:G_redef33} \\
  \partial_\zeta \Gamma (\zeta, \zeta') & = \int\limits_{0}^{\zeta'} d
  \xi' \left[ \Gamma (\zeta, \xi') + 3 \, G (\xi') \right]
  , \label{e:Gam_redef3}
\end{align}
\end{subequations}
and we then introduce the Laplace transforms
\begin{subequations} \label{e:Mellin}
\begin{align}
  & \hspace{-0.175cm} \!G (\zeta) = \int \frac{d \omega}{2 \pi i} \,
  e^{\omega \, \zeta} \, G_\omega,\!  & \hspace{-0.15cm} &\Gamma
  (\zeta, \zeta') = \int \frac{d \omega}{2 \pi i} \, e^{\omega \,
    \zeta'} \, \Gamma_\omega (\zeta) ,
\\
& \hspace{-0.175cm} G_\omega = \int\limits_0^\infty d\zeta \, e^{-
  \omega \zeta} \, G(\zeta),\!\!  & \hspace{-0.15cm} &\Gamma_\omega
(\zeta) = \int\limits_0^\infty d\zeta' \, e^{-\omega \zeta'}
\Gamma(\zeta, \zeta').
\end{align}
\end{subequations}
Consider first the Laplace transform \eqref{e:Mellin} of
\eq{e:Gam_redef3},
\begin{align}\label{eq:second}
\partial_\zeta \Gamma_\omega (\zeta) = \frac{1}{\omega} \, \left[ \Gamma_\omega (\zeta) + 3 \, G_\omega \right].
\end{align}
This is just an ordinary differential equation in $\zeta$, with the
solution
 \begin{align}\label{eq:Gamma1}
\Gamma_\omega (\zeta) + 3 \, G_\omega = e^{\frac{\zeta}{\omega}} \, 
\left[ \Gamma_\omega (0) + 3 \, G_\omega \right] ;
\end{align}
substituting \eqref{eq:Gamma1} back into \eqref{e:Mellin} then gives
\begin{align} \label{eq:Gamma2}
\hspace{-0.25cm}
\Gamma (\zeta, \zeta') = \int \frac{d \omega}{2 \pi i} \, e^{\omega \, \zeta'} \, 
\left\{ e^{\frac{\zeta}{\omega}} \, \left[ \Gamma_\omega (0) + 3 \, G_\omega \right] 
- 3 \, G_\omega \right\} ,
\end{align}
or, equivalently, 
\begin{align}\label{eq:Gamma+3G}
  \hspace{-0.25cm} \Gamma (\zeta, \zeta') + 3 \, G (\zeta') = \int
  \frac{d \omega}{2 \pi i} \, e^{\omega \, \zeta' +
    \frac{\zeta}{\omega}} \, \left[ \Gamma_\omega (0) + 3 \, G_\omega
  \right] .
\end{align}
Using the second boundary condition in \eqref{init}, \eq{eq:Gamma+3G}
then fixes $G$, giving the general solution for $G$ and $\Gamma$ as
\begin{subequations}\label{GGamma}
\begin{align}
  G (\zeta) &= \frac{1}{4} \, \int \frac{d \omega}{2 \pi i} \,
  e^{(\omega + \frac{1}{\omega}) \, \zeta} \, H_\omega
  , \label{GGamma1}
\\
\Gamma (\zeta, \zeta') &= \int \frac{d \omega}{2 \pi i} \, e^{\omega
  \, \zeta' + \frac{\zeta}{\omega}} \, H_\omega
\notag \\ &
- \frac{3}{4} \, \int \frac{d \omega}{2 \pi i} \, e^{(\omega +
  \frac{1}{\omega}) \, \zeta'} \, H_\omega ,
\end{align}
\end{subequations}
where we have introduced the unknown function $H_\omega$ as
\begin{align}\label{Hdef}
H_\omega \equiv \Gamma_\omega (0) + 3 \, G_\omega .
\end{align}

It is useful to observe that, upon substituting \eq{eq:Gamma2} back
into \eq{e:Gam_redef3}, the consistency of the solution requires that
\begin{align}\label{zero}
  \int \frac{d \omega}{2 \pi i} \, e^{\frac{\zeta}{\omega}} \,
  \frac{1}{\omega} \, H_\omega =0.
\end{align}
Indeed, the $\omega$ contour in the Bromwich integral \eqref{e:Mellin}
runs parallel to the imaginary axis and to the right of all the poles
of the integrand. Because the extra factor of $1/\omega$ in the
integrand of \eq{zero} provides sufficient convergence at infinity, we
can close the contour in the right half-plane, getting zero and
confirming \eq{zero}.

Finally, we can impose a further constraint on our results in
Eqs.~\eqref{GGamma} by requiring them to also satisfy
\eq{e:G_redef33}. Plugging Eqs.~\eqref{GGamma} into \eq{e:G_redef33}
and employing \eqref{zero} gives the constraint
\begin{align}\label{zero1}
  \int \frac{d \omega}{2 \pi i} \, e^{\omega \, \zeta +
    \frac{\zeta}{\omega}} \, \left( \omega - \frac{3}{\omega} \right)
  \, H_\omega = 0 .
\end{align}
It is convenient to define $f_\omega$ such that
%
\begin{align} \label{e:fdef}
%
%
H_\omega &= \left( \frac{\omega}{\omega^2 - 3} \right) f_\omega
\end{align}
%
and expand $f_\omega$ in a Laurent series:
\begin{align} \label{e:Laurent}
f_\omega = \sum_{n=-\infty}^\infty \, c_n \, \omega^n .
\end{align}
%
After expanding both $f_\omega$ with \eqref{e:Laurent} and $e^{\zeta /
  \omega}$ in their respective series, we pick up the enclosed
residues at $\omega = 0$ and obtain the constraint
\begin{align}\label{zero3}
  0 &= \int \frac{d \omega}{2 \pi i} \, e^{\omega \, \zeta +
    \frac{\zeta}{\omega}} \, f_\omega
\notag \\ &=
\sum_{n = -\infty}^{-1} c_{n} \, I_{-n-1} (2\zeta) + \sum_{n=0}^\infty
c_{n} \, I_{n+1} (2\zeta)
\notag \\ &=
c_{-1} I_0 (2\zeta) + \sum_{n=1}^\infty (c_{n-1} + c_{-n-1}) I_n
(2\zeta) .
\end{align}
Thus, we obtain $c_{-1} = 0$ and $c_n = - c_{-n-2}$ for $n \geq 0$,
such that
\begin{align} \label{e:fsoln}
f_\omega = \sum_{n=0}^\infty c_n \left( \omega^n - \frac{1}{\omega^{n+2}} \right) .
\end{align}
However, we know that $f_\omega$ cannot contain large positive powers
of $\omega$, or else it would affect convergence at infinity and
violate the consistency condition \eqref{zero}.  Substituting
\eqref{e:fdef} into \eqref{zero} gives
\begin{align}
  0 &= \int\frac{d\omega}{2\pi i} \, e^{\frac{\zeta}{\omega}} \,
  \frac{1}{\omega^2 - 3} f_\omega.
\end{align}
Taking $\zeta = 0$ for simplicity and using \eqref{e:fsoln}, we have
\begin{align}
0 &= \sum_{n = 0}^{\infty} c_n \int\frac{d\omega}{2\pi i} \, \frac{1}{\omega^2 - 3} 
\left( \omega^n - \frac{1}{\omega^{n + 2}} \right)
\notag \\ &=
\sum_{n = 1}^{\infty} c_n \int\frac{d\omega}{2\pi i} \, \frac{\omega^n}{\omega^2 - 3} ,
\end{align}
where for all sufficiently convergent integrals, we have closed the
contour in the right-half plane and obtained zero.  Therefore, the
consistency condition \eqref{zero} implies that $c_n =0$ for $n \ge
1$, such that
%
\begin{align} \label{H}
%
%
H_\omega &= c_0 \frac{\omega^2 - 1}{\omega \, (\omega^2 - 3)}. 
\end{align}

The function \eqref{H} fixes the solution of the helicity evolution
equations, giving for $G$ in \eq{GGamma1}
\begin{align}\label{Gexact}
  G (\zeta) = \frac{c_0}{4} \, \int \frac{d \omega}{2 \pi i} \,
  e^{\omega \, \zeta + \frac{\zeta}{\omega}} \, \frac{\omega^2 -
    1}{\omega \, (\omega^2 - 3)}.
\end{align}
Using the first boundary condition in \eqref{init} at $\zeta = 0$
fixes the coefficient to $c_0 =4$, after closing the contour in the
left half-plane and collecting the residues at $\omega = 0 , \pm
\sqrt{3}$.  Therefore, the complete asymptotic solution of the
large-$N_c$ helicity evolution equations is given by
\begin{subequations}\label{GGGexact_final}
\begin{align} \label{Gexact_final}
  G (\zeta) & = \int \frac{d \omega}{2 \pi i} \, e^{\omega \, \zeta +
    \frac{\zeta}{\omega}} \, \frac{\omega^2 - 1}{\omega \, (\omega^2 -
    3)} ,
\\
\Gamma (\zeta, \zeta') & = 4 \int \frac{d \omega}{2 \pi i} \,
e^{\omega \, \zeta' + \frac{\zeta}{\omega}} \, \frac{\omega^2 -
  1}{\omega \, (\omega^2 - 3)}
\notag \\ & - 
3 \int \frac{d \omega}{2 \pi i} \, e^{\omega \, \zeta' +
  \frac{\zeta'}{\omega}} \, \frac{\omega^2 - 1}{\omega \, (\omega^2 -
  3)}.
\end{align}
\end{subequations}
The high-energy $\!\!$/$\!\!$ small-$x$ asymptotics of \eq{Gexact_final},
corresponding to $\zeta \sim \zeta' \gg 1$, are given by the
right-most pole of the integrand at $\omega = + \sqrt{3}$.  Keeping
the contribution to \eqref{GGGexact_final} from this pole only, we
obtain the final result
\begin{subequations} \label{e:Gasms}
\begin{align} \label{e:Gasm}
G(\zeta) &\approx \frac{1}{3} e^{\frac{4}{\sqrt{3}} \zeta} 
\\
\Gamma(\zeta, \zeta') &\approx \frac{1}{3} e^{\frac{4}{\sqrt{3}}
  \zeta'} \left( 4 e^{\frac{\zeta - \zeta'}{\sqrt{3}}} - 3 \right)
\notag \\ \label{e:GamRatio} &=
G(\zeta') \left( 4 e^{\frac{\zeta - \zeta'}{\sqrt{3}}} - 3 \right) .
\end{align}
\end{subequations}
The asymptotic form of $G \sim e^{\frac{4}{\sqrt{3}} \zeta} \sim (z
s)^{\alpha_h}$ in \eqref{e:Gasm} gives the analytic expression for the
helicity intercept
\begin{align} \label{e:alphah3}
  \alpha_{h} &= \frac{4}{\sqrt{3}} \, \sqrt{\frac{\as N_c} {2\pi}}\,
  \approx 2.3094 \, \sqrt{\frac{\as N_c} {2\pi}} ,
\end{align}
in complete agreement with the numerical solution $\alpha_h \approx
2.31 \, \sqrt{\frac{\as N_c} {2\pi}}$ of \cite{Kovchegov:2016weo}!

\begin{figure}[ht]
\includegraphics[width=0.47\textwidth]{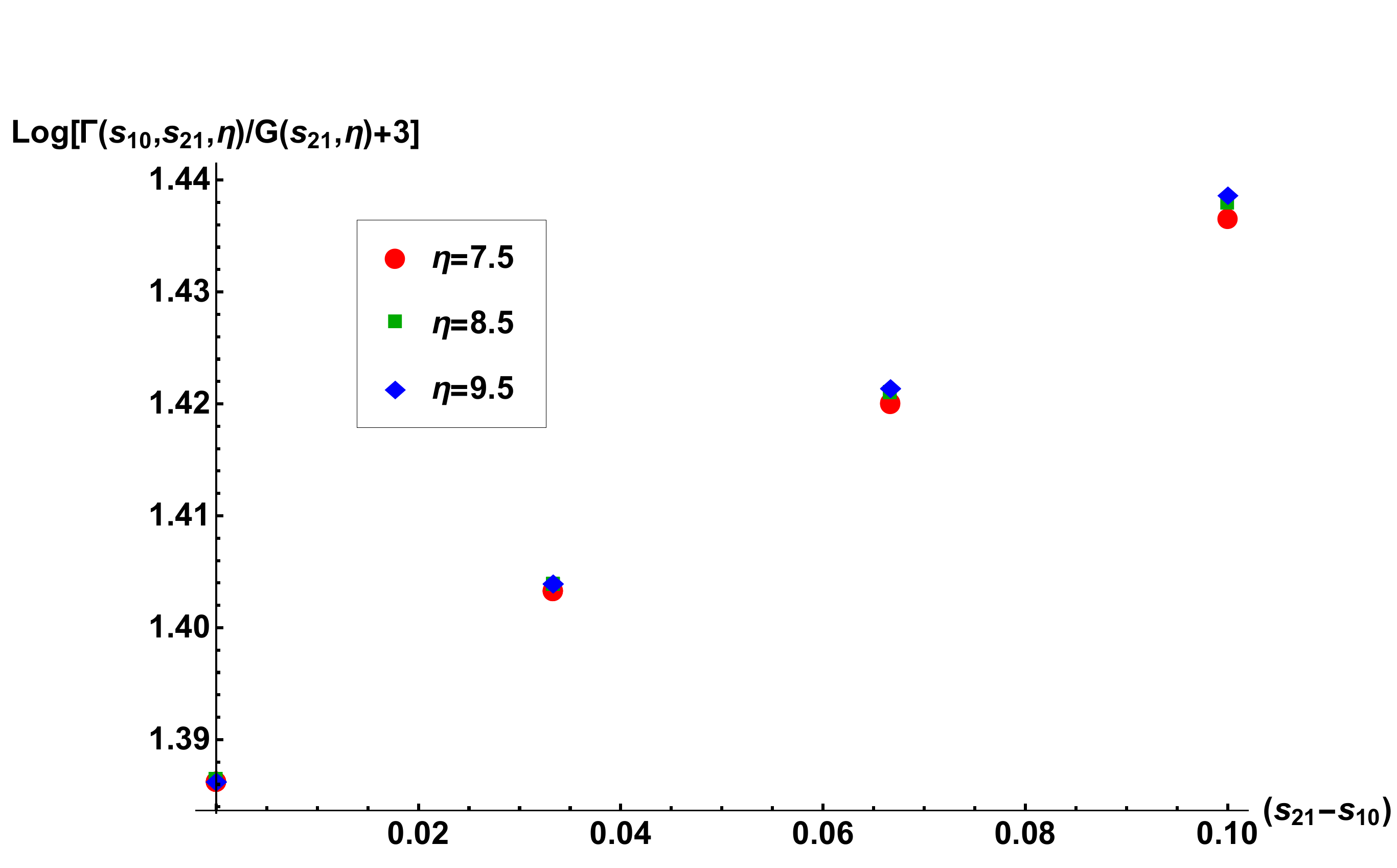}\\
\includegraphics[width=0.48\textwidth]{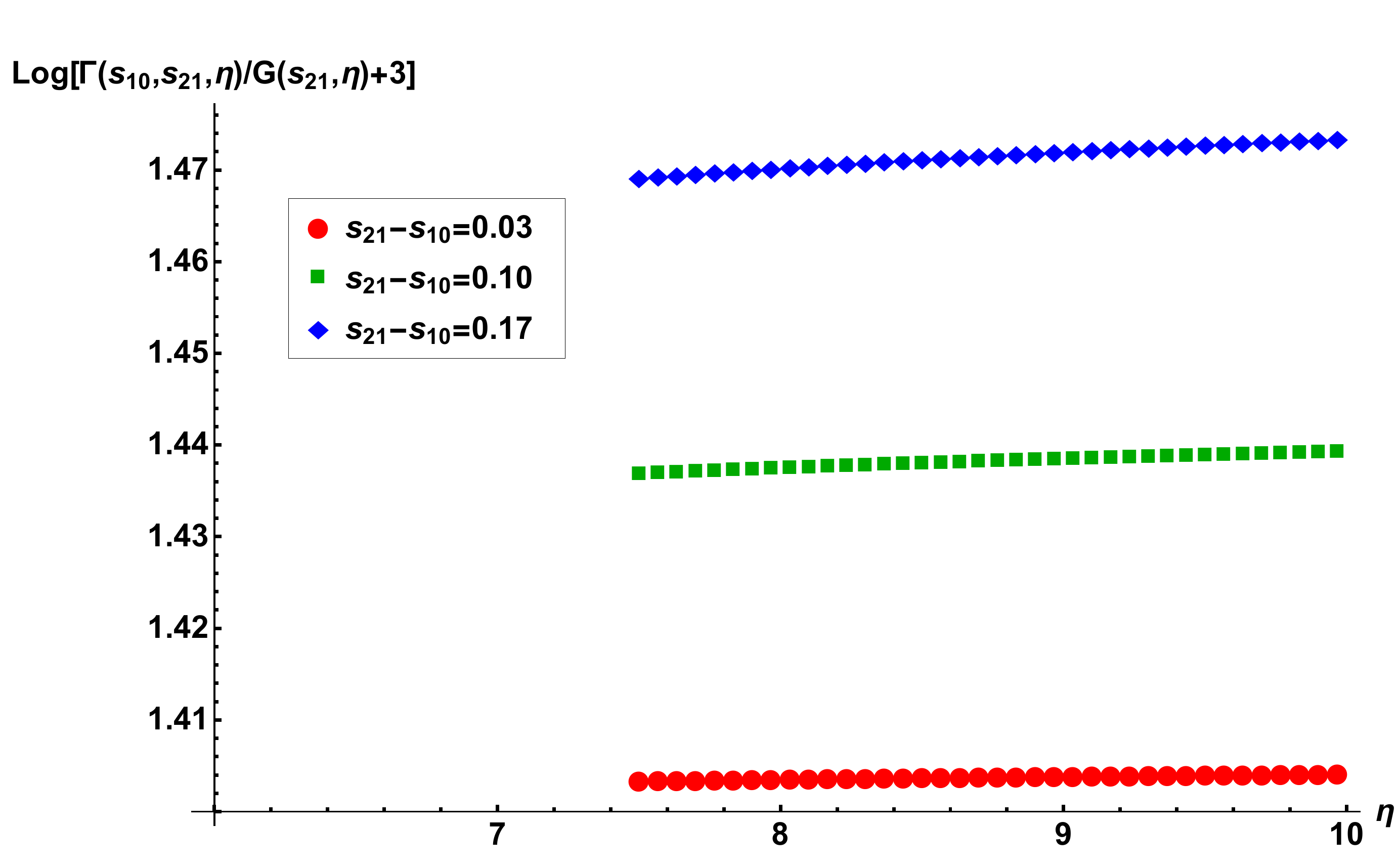}
\caption{ \label{f:num2} Plot of the scaling ratio \eqref{e:GamRatio2}
  in the numerical solution of \cite{Kovchegov:2016weo} as a function
  of $s_{21} - s_{10}$ for fixed $\eta$ (top panel) and as a function
  of $\eta$ for fixed $s_{21} - s_{10}$ (bottom panel).  The grid
  parameters are $\eta_{max} = 10 , \Delta\eta = 0.03$.  }
\end{figure}

Finally, we note that \eqref{e:GamRatio} makes a useful prediction for
the form of $\Gamma$ which can be straightforwardly tested against the
existing numerical solution of \cite{Kovchegov:2016weo}.  In the units
\eqref{e:logunits} used in the numerics, our analytic solution
predicts that the ratio of $\Gamma$ to $G$ should scale as
\begin{align} \label{e:GamRatio2}
  \ln\left[ \frac{\Gamma(s_{10} , s_{21} , \eta)}{G(s_{21} , \eta)} +
    3 \right] = \ln4 + \frac{1}{\sqrt{3}} (s_{21} - s_{10}) .
\end{align}
This ratio, calculated in the numerical solution of
\cite{Kovchegov:2016weo}, is plotted in Fig.~\ref{f:num2}, where we
see excellent agreement with the features of \eqref{e:GamRatio2}.
Qualitatively, no dependence of this ratio on $\eta$ is seen, and the
dependence on $s_{21} - s_{10}$ is linear.  And even though we have
not performed a detailed extrapolation from the discretized numerics
to the continuum, we even see
significant quantitative agreement with \eqref{e:GamRatio2}: the
vertical intercept (in the top panel of Fig.~\ref{f:num2}) of $1.386$
agrees fantastically with the expected $\ln 4 \approx 1.3863$, and the
slope of $\approx 0.52$ is within $10\%$ of the expected
$\frac{1}{\sqrt{3}} \approx 0.577$.  Indeed, if we perform a general
fit of $\ln [ \frac{\Gamma(s_{10} , s_{21} , \eta)}{G(s_{21}, \eta)} +
3]$ for $0 \leq s_{10} \leq s_{21} \leq 0.10$ and $7.5 \leq \eta \leq
10$ to a function of the form $a s_{21} + b s_{10} + c \eta + d$, we
find $a \approx - b \approx \frac{1}{\sqrt{3}}$ (within $10\%$
accuracy) and $c \approx 0$, $d
\approx \ln 4$ (with much greater accuracy).  This preferred 
functional form is in excellent agreement with our analytic calculation
\eqref{e:GamRatio2}.  We also note that the numerics in \cite{Kovchegov:2016weo} used scaling-violating initial conditions, so that the agreement
seen here validates our claim of negligible dependence on the initial
conditions. Thus, we can conclude with confidence that our
analytic solution \eqref{e:Gasm} and helicity intercept
\eqref{e:alphah3} are the correct generalization of the numerical
calculation in~\cite{Kovchegov:2016weo}.

\section{Conclusions}

In this Letter, we have derived an analytic solution to the
large-$N_c$ helicity evolution equations \eqref{e:helevol} in the
high-energy $\!\!$/$\!\!$ small-$x$ asymptotics.  The central results are the
solutions \eqref{e:Gasms} for the polarized dipole amplitude $G$ and
the auxiliary neighbor dipole amplitude $\Gamma$, leading to the
analytic expression for the helicity intercept \eqref{e:alphah3}.  The
key assumption which made such an analytic solution possible was the
observation of emergent scaling behavior \eqref{scaling} as seen in
the previous numerical solution of \cite{Kovchegov:2016weo}
(Fig.~\ref{f:num1}).  We have checked our analytic results by
comparing the predicted behavior of the auxiliary neighbor dipole
amplitude $\Gamma$ in \eq{e:GamRatio2} with the numerical solution in
Fig.~\ref{f:num2}, finding excellent agreement.

Unfortunately, it is not clear whether the techniques used here can be
extended to obtain an analytic solution of the helicity evolution
equations in the large-$N_c \, \& \, N_f$ limit \cite{Kovchegov:2015pbl,
  Kovchegov:2016zex}.  The addition of quark loops to the evolution
kernel introduces terms which explicitly break the scaling property
\eqref{scaling}, similar to what was found for the Reggeon
\cite{Itakura:2003jp} (see also \cite{Iancu:2015vea}).  
Therefore, we set aside the question of
generalizing this approach to the large-$N_c \, \& \, N_f$ limit as a
separate project, which we leave for future work.


\section*{Acknowledgments}

The authors are greatly indebted to Edmond Iancu for
his suggestion to look for a scaling solution to the helicity
evolution equations and to Bin Wu for several helpful
discussions of the involved integrals.

This material is based upon work supported by the U.S. Department of
Energy, Office of Science, Office of Nuclear Physics under Award
Number DE-SC0004286 (YK), within the framework of the TMD Topical
Collaboration (DP), and DOE Contract No. DE-AC52-06NA25396 (MS).  MS
received additional support from the U.S. Department of Energy, Office
of Science under the DOE Early Career Program. \\

\end{document}